\documentclass[runningheads]{llncs}

\usepackage[utf8]{inputenc}
\usepackage{graphicx}
\usepackage[table]{xcolor}
\usepackage[ruled,vlined]{algorithm2e}
\usepackage[colorlinks,citecolor=red]{hyperref}

\begin{document}

\title{Towards Modularity Optimization Using Reinforcement Learning to Community Detection in Dynamic Social Networks}
\titlerunning{Towards Modularity Optimization Using RL to Community Detection}

\author{Aurélio Ribeiro Costa\inst{1}\orcidID{0000-0001-7352-1454} 
}
\authorrunning{Costa, A.R.}

\institute{Computer Science Department, University of Brasília, Federal District, Brazil \\ \email{aurelio.ribeiro@aluno.unb.br}\\
\url{http://www.cic.unb.br}
}


\maketitle
\begin{abstract}
The identification of community structure in social network is an important problem tackled in literature of network analysis. There are many solutions to this problem using a static scenario, when facing a dynamic scenarios some solutions may be adapted but others simply do not fit, moreover when considering the demand to analyze constantly growing networks. In this context, we propose an approach to the problem of community detection in dynamic networks based on reinforcement learning strategy to deal with changes on big networks using a local optimization on modularity of the changed entities. An experiment using synthetic and real-world dynamic network data shows results comparable to static scenarios.

\end{abstract}

\section{Introduction}

Many contexts of real-world can be modeled as a network, relations between people~\cite{wasserman1994social,sharma2017community,tang2008arnetminer}, proteins interactions~\cite{amitai2004network}, fraud detection~\cite{vsubelj2011expert} and supply chains~\cite{hearnshaw2013complex} are some examples of contexts that can be modeled as networks. Carley cites in~\cite{carley2003dynamic} complex system as an entity modeled as a network. These networks can be classified by many criteria as dynamics or static, directed, regular, complex, or random. 

Network analysis is a tool that tries to evaluate the target networks in different aspects like node classification, link prediction, or community detection (CD). Each of these aspects can be analyzed alone or in an integrated way. Challenges and opportunities to classic methods of CD, such as spectral clustering and statistical inference, are being replaced by deep learning techniques with an increased capacity to handle high-dimensional graph data with an impressive performance as presented in~\cite{Liu2020}.

In the context of CD, to extract communities from a given network, one typically chooses a scoring function (e.g., modularity) that quantifies the intuition that communities correspond to densely linked sets of nodes. Then, one applies a procedure to find sets of nodes with a high value of the scoring function. This procedure can take two directions: agglomerating, when a set of nodes are merged, or dividing when edges are removed from the network and the scoring function is recalculated. Identifying such communities in networks has proven to be a challenging task due to some reasons:
\begin{itemize}
\item there exist multiple structural definitions of network communities~\cite{radicchi2004defining}; 
\item even if we would agree on a single common structural definition (i.e., a single scoring function), the formalization of CD leads to NP-hard problems~\cite{schaeffer2007graph};
\item the lack of reliable ground-truth makes evaluation extremely difficult~\cite{yang2015defining};
\item the growing size of networks used to identify such communities. 
\end{itemize}

The search for a unified solution for the problem of CD is indeed hard. Nevertheless, there are many solutions for the problem of CD when considering some constraints as in analyzing static networks where we can employ classical methods like leading eigenvector~\cite{newman2006finding} or random walks~\cite{pons2005computing}.

In this challenging scenario, the hypothesis being hold in this work is the application of reinforcement learning (RL) as an adequate approach to optimize the modularity of CD solutions applied to dynamic social networks. In this work, we present an RL approach to CD in dynamic networks that is comparable in results to the approach presented in~\cite{Cordeiro2016}, using the modularity density score function.

The rest of the work presents in Section~\ref{sec:related} related work, in Section~\ref{sec:pre} preliminary concepts, in Section~\ref{sec:sol} the proposed solution, in Section~\ref{sec:exp} some experiments with results, and in Section~\ref{sec:conc} conclusion and directions for future work.

\section{Related Work}
\label{sec:related}

This section presents works related to CD approaches. The work of Valejo et al.~\cite{Valejo2014} implements a multilevel CD  algorithm with changes to tackle the scenario with overlapping communities since the original multilevel approach only deals with disjoint communities, we use this work to compare our performance on overlapping CD. 

Cordeiro et al.~\cite{Cordeiro2016} present a CD technique for dynamic networks that maintains the community structure always up-to-date following the addition or removal of nodes and edges. The proposed algorithm s a modification of the original Louvain method where dynamically added and removed nodes and edges only affect their related communities. In each iteration, the algorithm maintains unchanged all the communities that were not affected by modifications to the network. By reusing community structure obtained by previous iterations, the local modularity optimization step operates in smaller networks where only affected communities are disbanded to their origin. The stability of communities is also an improvement over
the original algorithm. Given that only parts of the network change during iterations, the non-determinism of the algorithm will have reduced effect on the community assignment. Most node community assignments remain unchanged between snapshots, providing better community stability than its static counterpart. This work is used as a reference for our performance approach. 

Martins et al.~\cite{Martins2020} define an improvement to the particle competition method to CD enabling this method to deal with unbalanced communities. It is an important and common aspect of complex networks with many nodes. This work inspired us to search for a solution with low time complexity.

Following the trail of machine learning applied to CD, the studies undertaken by~\cite{paim2020detecting,saghiri2019random} use directly RL to deal with CD. Paim et al.~\cite{paim2020detecting} use the classical approach of modularity maximization computed through a Multi-Agent RL (MARL) approach. Saghiri et al.~\cite{saghiri2019random} also use a classical approach to CD named \emph{random walks}. The goal of the study is to aggregate an intelligent model of \emph{random walks} as a problem-solving method. However, none of these studies are applied to dynamic networks.

The study conducted by~\cite{molokwu2020node} applies a convolutional neural network (CNN) to extract facts from a social network and proceed with node classification and CD. However, when conducting the CD they treat network clustering in a simplified way only employing node similarity to identify communities and not consider relations between nodes.

Zhang et al.~\cite{zhang2020seal} propose a semi-supervised solution named SEAL (\emph{Seed Expansion with Generative Adversarial Learning}) based on a graph neural network (GNN) acting as the discriminator module. Their solution finds communities considering network topology and node attributes.

Table~\ref{tab:relwork} summarizes the related work comparing the main attributes like method employed to CD and network used to validate the method. The column ``CD Method" shows some methods named traditional. These methods receive this name to apply only analytical techniques while others apply AI techniques. Note that this work is the only one that applies RL for CD in dynamic networks.

\rowcolors{2}{gray!25}{white}
\begin{table}[h!]
    \centering
    \caption{Related work overview.}
    \begin{tabular}{l|p{3.5cm}|p{4.5cm}}
    \rowcolor{gray!50}
    Work & CD Method & Network \\ 
    Valejo et al., 2014~\cite{Valejo2014}& traditional - multilevel overlapping communities  &Facebook
(social network) and Yeast (biological network)\\
    Cordeiro et al.,  2016~\cite{Cordeiro2016} & traditional - modularity maximization&high-energy physics theory citation network\\
    Martins \& Zhao, 2020~\cite{Martins2020} & traditional -  random walks&com-DBLP, com-Amazon, com-Youtube\\
    Paim et al., 2020~\cite{paim2020detecting}& artificial intelligence - MARL & Zachary’s Karate Club, Bottlenose Dolphins, Kreb’s Political Books, American College Football, Email-Eu-core and Reactome\\
    Molokwu et al., 2020~\cite{molokwu2020node}& artificial intelligence - CNN & Cora, CiteSeer, Facebook Page2Page, PubMed Diabetes, Internet Industry, Terrorist Relation\\
    This work & artificial intelligence - RL & com-DBLP, com-Amazon, com-Youtube\\
    \end{tabular}
    \label{tab:relwork}
\end{table}

\section{Preliminaries}
\label{sec:pre}

This section presents some fundamental aspects with concepts and artificial intelligence techniques related to CD being presented in Sections~\ref{cdconcepts} and~\ref{aitech}, respectively.

\subsection{CD concepts}
\label{cdconcepts}

\begin{figure}[h]
    \centering
    \includegraphics[width=0.5\textwidth,height=5cm]{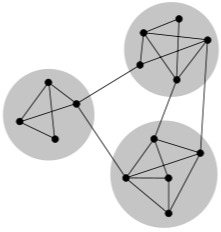}
    \caption{A network with three disjoint communities.}
    \label{fig:community}
\end{figure}

In a graph $\mathcal{G} = (\mathcal{V},\mathcal{E})$ with $\mathcal{V}= \{v_1, v_2, ...v_n\}$, $\mathcal{E} = \{e_1,e_2, ..., e_m\}$, $|\mathcal{V}|=n$ and $|\mathcal{E}|=m$, a community $\mathcal{C}$ is typically defined by sets of nodes  densely interconnected which are sparsely connected with the rest of the nodes. Finding communities within a graph helps unveil the internal organization of a graph, and can also be used to characterize the entities that compose it (e.g., groups of people with shared interests, products with common properties, etc.)~\cite{pares2017fluid}. In the scenario with a dynamic network, we can deal the dynamic network $\mathcal{G} = \{\mathcal{G}_0,\mathcal{G}_1,\mathcal{G}_2, ..., \mathcal{G}_n\}$ as a set of snapshots in time.  Figure~\ref{fig:community} shows a network highlighting three disjoint communities (i.e. non overlapping communities).

A common metric of CD-quality is the modularity introduced in~\cite{newman2006modularity} and represented by Equation~\ref{eq:modu} as a generalization to deal with $c$ communities. Using this equation one can compute the score modularity of a community structure $\mathcal{C}$ in a network $\mathcal{G}$, $Q$ is defined as the fraction of edges within communities minus the expected  fraction in a corresponding random network that servers as a null model, 
\begin{equation}
Q = \sum_{c \in C}\left[\frac{m_c}{m} - \frac{2m_c+e_c}{2m}\right]\delta(c_u,c_v), 
\label{eq:modu}
\end{equation}
where $m_c$ is the number of edges in community $c$, $e_c$ is the number of external edges of $c$, and $m$ is the total number of edges in the network. The partition that maximizes $Q$ is considered as the one that corresponds to the community structure.

Chen et al.~\cite{chen2018network} cite that despite its popularity, the metric Q has drawbacks. Perhaps the most notable is that by maximizing Q one may not detect communities that contain fewer links than $m_c \sim \sqrt{2m}.$ This is known as the resolution limit problem. Perhaps the most promising approach is to use a new metric called modularity density $Q _{ds}$ to quantify community structure defined as:
\begin{equation}
    Q_{ds} = \sum_{c \in C}\left[\frac{m_c}{m}p_c - \left(\frac{2m_c + c_e}{2m}p_c\right)^2 - \sum_{c' \neq c}\frac{m_{cc'}}{2m}p_{cc'}\right],
\label{eq:modudensity}
\end{equation}
where $m_{cc'}$ is the number of edges between communities $c$ and $c'$, $n_c$ is the number of nodes in $c$, $p_c = \frac{2m_c}{n_c(n_c-1)]}$ is the density of links inside $c$, $p_{cc'} = \frac{m_{cc'}}{n_cn_{c'}} $ is the density of edges between $c$ and $c'$, and the other quantities are the same as in Equation~\ref{eq:modu}. Again, it is the partition that maximizes $Q_{ds}$ that corresponds to the community structure.

\subsection{AI techniques}
\label{aitech}

RL is a subfield of AI that explicitly considers the whole problem of a goal-directed agent interacting with an uncertain environment~\cite{sutton2018reinforcement}. In the general architecture of an RL model, there are two main elements the agent and the environment. The environment can be defined as the \textit{locus} where the agent operates, on the other hand, the agent is the element responsible for observing the environment and take actions that change the environment. The agent receives a reward for each action taken and its final objective is to maximize the accumulated reward. The actions taken by the agent follow a policy that, in this work, implements the control method State-Action-Reward-State-Action (SARSA), described in Equation \ref{eq:sarsa} where $\alpha$ and $\gamma$ are the learning rate and future reward discount, respectively, \emph{s} is a state, \emph{a} is an action and \emph{r} is the reward when applying a  in the state s. 

\begin{equation}
    Q(s_t,a_t) = Q(s_t,a_t) + \alpha(r_{t+1} + \gamma Q(s_{t+1},a_{t+1}) - Q(s_t,a_t))
    \label{eq:sarsa}
\end{equation}

There are many implementations of RL however this work adopt the Q-Learning approach. The experiment discussed in this paper uses four implementations of CD algorithms, each one using a different CD approach, but other implementations can be used with the proposed solution:
\begin{itemize}
\item the leading eigenvector algorithm was proposed by Newman in \cite{newman2006finding}. The basic idea is to find the eigenvalue of the largest magnitude and its eigenvector for a modularity matrix representing the network. It is the same as finding the community structure that maximizes the modularity score for that network.
\item the walktrap algorithm, proposed by~\cite{pons2005computing}, explores the intuition that random walks on a graph tend to get ``trapped” into densely connected parts corresponding to communities. Walktrap uses a measurement of the structural similarity between vertices and communities. Thus, defining a distance between those entities. The authors point that this method has a time complexity of $\mathcal{O}(n^2log(n))$ being memory costly once many walks (sequence of nodes) are kept in memory. 
\item the label propagation algorithm, proposed by \cite{raghavan2007near}, in this method every node is initialized with a unique label and at every step, each node adopts the label that most of its neighbors currently have. In this iterative process, densely connected groups of nodes form a consensus on a unique label to form communities.
\item the multilevel algorithm, proposed by \cite{blondel2008fast}, is a bottom-up algorithm where initially every vertex belongs to a separate community, and vertices are moved between communities iteratively in a way that maximizes the vertices' local contribution to the overall modularity score. When a consensus is reached (i.e. no single move would increase the modularity score), every community in the original graph is shrank to a single vertex (while keeping the total weight of the adjacent edges) and the process continues on the next level. The algorithm stops when it is not possible to increase the modularity anymore after shrinking the communities to vertices.
\end{itemize}

\section{Proposed Solution}
\label{sec:sol}

The proposed solution consists of a RL method to find the optimal combination of algorithm and parameters set of the CD algorithms applied to a dynamic network. This approach is based on classical solutions of CD as leading eigenvector~\cite{newman2006modularity}, walktrap~\cite{pons2005computing}, label propagation~\cite{raghavan2007near} and multilevel~\cite{blondel2008fast}. The RL method is characterized as model-free, value-based and on-policy using SARSA as update control method and $\epsilon$-greedy approach to choose action. The RL method uses Q-Learning to store the combination of algorithm and parameters and the modularity score as reward in its Q matrix of knowledge.

Algorithm~\ref{lst:algo} implements the RL agent's main loop. Algorithm~\ref{lst:pickalgo} implements the policy and receives the argument \emph{same} that returns the previous CD method with changes in its parameters. $Q(G,c)$ is the CD quality metric adopted, to make a viable comparison with \cite{Cordeiro2016} we adopted the modularity density. The discount factor $\gamma$ gives a margin to not change the CD method very frequently and gives space to improve the reward by changing only the method's parameter.

\begin{algorithm}[h!]
\SetAlgoLined
\KwIn{G \% Social network\\ max\_episodies \% number of iterations\\ $\alpha$ \% learning rate\\ $\gamma$ \% discount factor\\ $\epsilon$ \% greedy factor}
$q_{max} \leftarrow argmaxQ(G)$\;
C $\leftarrow null$ \;
r $\leftarrow$ 0\;
done $\leftarrow$ False\;
Q $\leftarrow$ 0\;
\For{$t < max\_episodies$}{
    action,parameters = $\leftarrow$ improve\_modularity\_policy(C, r, $\epsilon$)\;
  \While{not done}{
    $C \leftarrow action(G, parameters)$\;
    $r \leftarrow Q_{ds}(G,C)$\;
    
    action2,parameters2 $\leftarrow$ improve\_modularity\_policy(C, r, $\epsilon$)\;
    
    predict $\leftarrow$ Q[action,parameters]\;
    target $\leftarrow$ r + $\gamma$Q[action2,parameters2]\;
    
    Q[action,parameters] = Q[action,parameters] + $\alpha(target-predict)$\;
    
    action $\leftarrow$ action2\;
    parameter $\leftarrow$ parameter2\;
    
    \If{$r = q_{max}$} {
      done $\leftarrow$ True\;
    }
  }
}
\label{lst:algo}
\caption{Community detection agent.}
\end{algorithm}

Algorithm~\ref{lst:pickalgo} implements the RL $\epsilon$-greedy policy. It returns the action that can be a random algorithm with a random set of parameters or the the action that maximize the reward of state \emph{c} according to the parameter $\epsilon$.
\begin{algorithm}[h!]
\SetAlgoLined
\KwIn{\\
c \% community structure\\
r \% reward\\
$\epsilon$ \% greedy offset\\
}
\KwOut{A community detection algorithm and its parameters}
action $\leftarrow$ null\;
\eIf{$random(0,1) < \epsilon$}{
    action $\leftarrow$ (random\_action, random\_parameters)\;}
{
  action $\leftarrow$ argmax(Q[c, ])\;
}

\If{$r > 0$}{
update\_parameters\;
}

return action\;
  
\label{lst:pickalgo}
\caption{Improve modularity policy.}
\end{algorithm}

\begin{figure}
    \centering
    \includegraphics[width=0.8\textwidth]{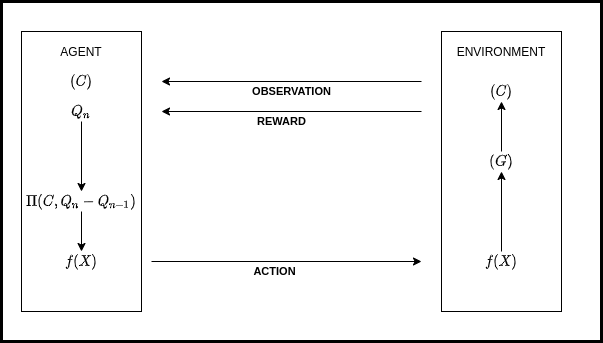}
    \caption{The conceptual  architecture.}
    \label{fig:rl2cdarch}
\end{figure}

Figure~\ref{fig:rl2cdarch} presents the conceptual architecture highlighting the interaction between agent, environment, and internal aspects of these entities, where:
\begin{itemize}
    \item $G$ is the network;
    \item $f(\cdot)$ is a CD method (CDM);
    \item $X$ is the CDM parameters;
    \item $C$ is the community structure;
    \item $Q$ is the modularity score; 
    \item $Q_{n}$ is $Q$ in episode $n$; and
    \item $\Pi(\cdot)$ is the policy.
\end{itemize}

\section{Experiment and Discussion}
\label{sec:exp}

An experiment was undertaken to validate the accuracy of our solution. We choose the algorithms cited in Section~\ref{sec:sol}, each one implementing one kind of CD approach, to implement the RL's action space. The experiment used a synthetic network generated using the Erdos Renyi model to validate the RL implementation and a real-world dataset of High-energy physics theory citation network available in Snap Project\footnote{\url{http://snap.stanford.edu/data/cit-HepTh.html}}, Table~\ref{tab:datasets} describes the real-world dataset statistics.

\begin{table}[h!]
    \centering
    \caption{High-energy physics theory citation network dataset statistics.}
    \begin{tabular}{l||r}
    \rowcolor{gray!50}
        Feature & Value\\
        Nodes & 27,770\\
        Edges & 352,807\\
        Nodes in largest WCC & 27,400 (98.7\%)\\ 
        Edges in largest WCC & 352,542 (99.9\%)\\
        Nodes in largest SCC & 7,464 (26.9\%)\\
        Edges in largest SCC & 116,268 (33\%)\\
        Average clustering coefficient & 0.3120\\
        Number of triangles & 1,478,735\\
        Fraction of closed triangles & 0.04331\\
        Diameter (longest shortest path) & 13\\
        90-percentile effective diameter & 5.3\\
    \end{tabular}
    \label{tab:datasets}
\end{table}

The work of Sousa and Zhao~\cite{Sousa2014} presents an evaluation of CD solutions available in the iGraph package in a set of different scenarios. Inspired on this evaluation, the RL framework presented in this work uses the iGraph package as an implementation of different CD algorithms~\cite{Csardi2006}. To evaluate the quality of community structure we used the modularity score as described in Section~\ref{sec:pre}. We used the OpenAI Gym library to implement the RL' environment together with the iGraph to handle the network data.

\begin{figure}[h!]
    \centering
    \includegraphics[width=0.8\textwidth]{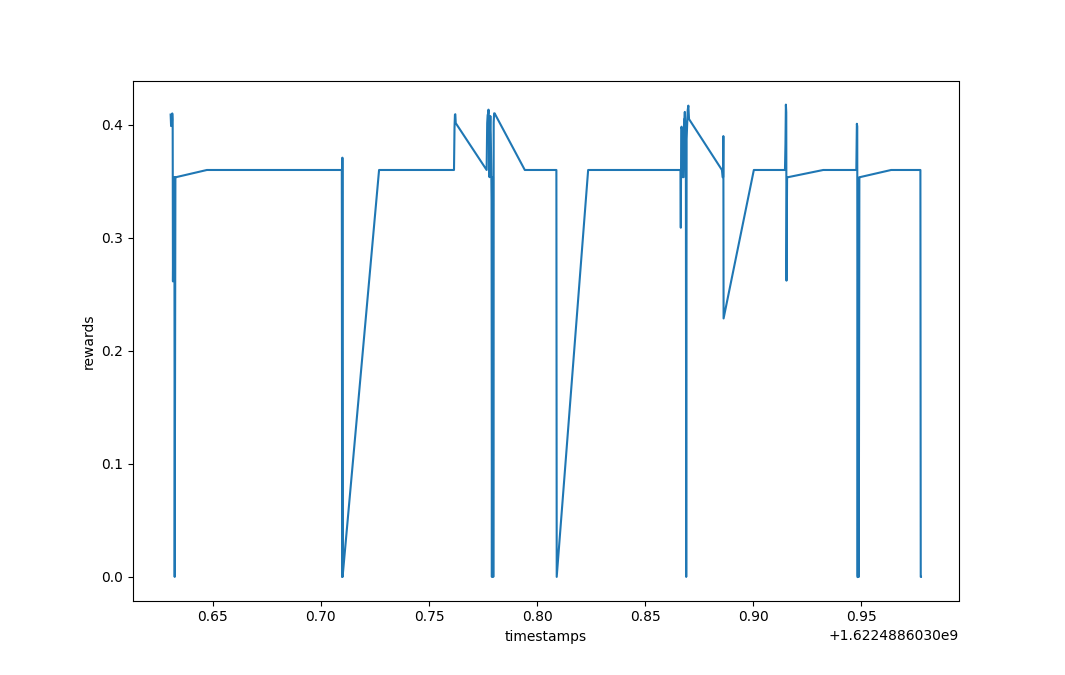}
    \caption{Null model for experiment.}
    \label{fig:nullmodel}
\end{figure}

\begin{figure}[h!]
    \centering
    \includegraphics[width=0.8\textwidth]{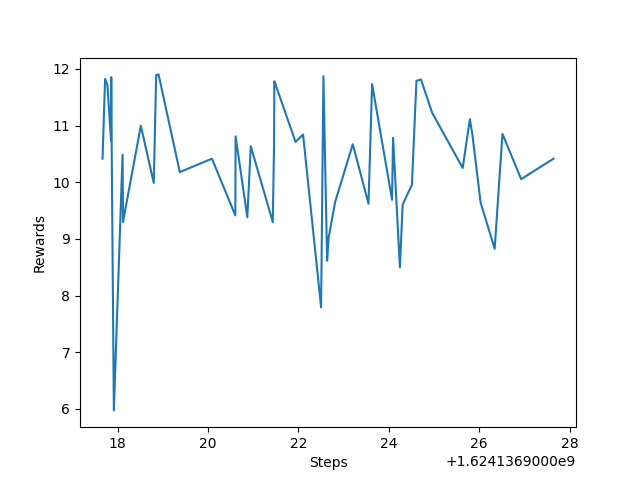}
    \caption{Accumulated reward evolution over episodes.}
    \label{fig:result1}
\end{figure}

A null model was used as the first iteration of our experiment, using $\epsilon = 1$ to make action selection entirely random. The policy selects only random actions and the result is shown in Figure~\ref{fig:nullmodel}. This null model was set as a baseline for comparison then we improve the policy to take actions based on a $\epsilon$-greedy approach to create a balance between exploration and exploitation depending on the value of $\epsilon$. The choice to use SARSA as an update method for our policy was mainly driven by the unpredictability of the future states on a dynamics network, . Figure \ref{fig:result1} shows the variation of modularity in time using the best RL configuration parameters, these parameters are described on Table \ref{tab:params}.

\begin{table}[htb]
    \centering
    \caption{Configuration parameters.}
    \begin{tabular}{c|r|c}
    \rowcolor{gray!50}
        Parameter & Value & Description \\
        episodes & 50 & Number of executions\\
        $\alpha$ & 0.8 & Learning rate\\
        $\gamma$ & 0.5 & Future reward discount\\
        $\epsilon$ & 0.2 & Update discount
    \end{tabular}
    \label{tab:params}
\end{table}

The execution of our RL implementation  with the High-energy physics theory citation network resulted in a plot shown in Figure \ref{fig:result1} where we can see a evolution of the accumulated reward over episodes. The results found by Cordeiro et al.~\cite{Cordeiro2016} were used as ground-truth and the comparison is present in Table \ref{tab:resultcompare}.

\begin{table}[htb]
    \centering
    \caption{Experiment comparison with ground-truth}
    \begin{tabular}{c|c|c}
    \rowcolor{gray!50}
         & Ground-truth & Our implementation  \\
         Average Modularity & 0.60396 & 0.6561 \\
         Execution time (s) & 145 &  6,530
    \end{tabular}
    
    \label{tab:resultcompare}
\end{table}

\section{Conclusion}
\label{sec:conc}

This work introduced a RL approach to the CD problem of dynamic networks what validates the hypothesis that  the application of reinforcement learning (RL) is an adequate approach to optimize the modularity of CD solutions applied to dynamic social networks. It was demonstrated that classical implementations of CD algorithms as leading eigenvector, multilevel, random walk and label propagation can be used as core components in action space of a RL. This approach highlights the flexibility of RL to incorporate classical solution to constrained problems to more general scenarios.

An experiment was conducted with real datasets and the evaluation shows that the results are comparable to the literature for CD using dynamic networks. 

As future work we intend to investigate other approaches of RL specifically that using the Actor-Critic architecture to improve the stability of reward. Another future work is to try other implementations that can deal with big networks efficiently.


\bibliographystyle{splncs03_unsrt}
\bibliography{bracis}
\end{document}